\documentclass[mathleft]{an}
\usepackage{graphicx}
\usepackage{times}
\usepackage{url}
\begin{document}
 \Pagespan{001}{}
 \Yearpublication{XXXX}%
 \Yearsubmission{2015}%
 \Month{XX}%
 \Volume{XXX}%
 \Issue{XX}%

 \title{On the angular distribution of IceCube high-energy events
       }
 \author{R. de la Fuente Marcos\thanks{Corresponding author:\email{rauldelafuentemarcos@gmail.com}}
         \and
         C. de la Fuente Marcos
        }
 \authorrunning{de la Fuente Marcos \& de la Fuente Marcos}
 \titlerunning{On the angular distribution of IceCube events}
 \institute{Apartado de Correos 3413, E-28080 Madrid, Spain
           }

 \received{2015 Jun 16}
 \accepted{2015 Aug 06}
 \publonline{2015 Sep 11}

 \keywords{Galaxy: center -- galaxies: clusters: general -- 
           methods: numerical -- methods: statistical -- neutrinos 
          }
 \abstract
   {The detection of high-energy astrophysical neutrinos of extraterrestrial
    origin by the IceCube neutrino observatory in Antarctica has opened a
    unique window to the cosmos that may help to probe both the distant
    Universe and our cosmic backyard. The arrival directions of these
    high-energy events have been interpreted as uniformly distributed on the
    celestial sphere. Here, we revisit the topic of the putative isotropic 
    angular distribution of these events applying Monte Carlo techniques to 
    investigate a possible anisotropy. A modest evidence for anisotropy is 
    found. An excess of events appears projected towards a section of the 
    Local Void, where the density of galaxies with radial velocities below 
    3000 km s$^{-1}$ is rather low, suggesting that this particular group of 
    somewhat clustered sources are located either very close to the Milky 
    Way or perhaps beyond 40 Mpc. The results of further analyses of the 
    subsample of southern hemisphere events favour an origin at cosmological 
    distances with the arrival directions of the events organized in a 
    fractal-like structure. Although a small fraction of closer sources is
    possible, remote hierarchical structures appear to be the main source of 
    these very energetic neutrinos. Some of the events may have their origin 
    at the IBEX ribbon.
   }

 \maketitle

 \section{Introduction}
    The IceCube neutrino observatory has announced the observation of very high-energy events that have been interpreted as the first robust 
    evidence of neutrinos from extraterrestrial origin (Aartsen et al. 2013a,b,c, 2014a; Halzen 2014) other than those detected from the 
    supernova 1987A (see, e.g., Arnett et al. 1989) or the Sun (see, e.g., Bellerive 2004; Bahcall 2005). A total of 37 events have been 
    found in data collected from May 2010 to April 2013, an all-sky search over a period of 988 days, and their deposited energies range 
    from $\sim$30 TeV to $\sim$2 PeV. Assuming an astrophysical origin for these detections, a number of possible sources have been 
    suggested (see, e.g., Murase et al. 2013; Anchordoqui et al. 2014a; Joshi et al. 2014; Ahlers \& Murase 2014). These include both 
    Galactic and extragalactic sources (Halzen 2014) as well as others more exotic and rare (see, e.g., Bhattacharya et al. 2010). 
    
    Among the proposed Galactic sources we find the centre of the Galaxy (see, e.g., Razzaque 2013a), its halo (see, e.g., Taylor et al. 
    2014) with the Fermi Bubbles (see, e.g., Lunardini et al. 2014), and supernova remnants (see, e.g., Padovani \& Resconi 2014). Suggested 
    extragalactic sources range from active galactic nuclei (AGN; Winter 2013; Krau\ss\ et al. 2014; Padovani \& Resconi 2014), radio 
    (Becker Tjus et al. 2014) or starburst galaxies (Anchordoqui et al. 2014b; Liu et al. 2014) to gamma-ray bursts (GRBs; see, e.g., 
    Razzaque 2013b). Decay of a particle much heavier than PeV can be another possible source (Ema et al. 2014). Dark matter annihilation 
    has also been suggested as the origin of the most energetic ones (Zavala 2014) although they can be explained satisfactorily within the 
    framework of the Standard Model of neutrino-nucleon interactions (Chen et al. 2014). 

    Whatever the source, the angular distribution of events is considered consistent with an isotropic arrival direction of neutrinos 
    (Aartsen et al. 2013c, 2014a) and the majority of events detected by IceCube are believed to be from cosmological distances or, less 
    likely, from the Galactic halo. In a recent update, Aartsen et al. (2014b) have found no evidence of neutrino emission from discrete or 
    extended sources in four years of IceCube data. However, and although any distribution of halo sources is expected to be isotropic, the 
    spatial distribution of galaxies (star-forming or quiescent) located at cosmological distances is far from uniform, being arranged in a 
    hierarchical fashion (see, e.g., Pietronero 1987).  

    Here, we revisit the topic of the possible isotropic angular distribution of IceCube high-energy events using a Monte Carlo approach. 
    This paper is organized as follows. The characteristics of the data used in our study are summarized in Sect. 2. Section 3 presents our 
    Monte Carlo-based angular distribution analysis; an area of the sky is identified as key to understand the most probable origin of the 
    events studied here. The atmospheric background is further explored in Sect. 4. Our results are discussed within the context of the 
    fractal structure of the Universe in Sect. 5. Conclusions are summarized in Sect. 6.

 \section{The data}
\begin{figure}
       \resizebox{\hsize}{!}{
          \includegraphics{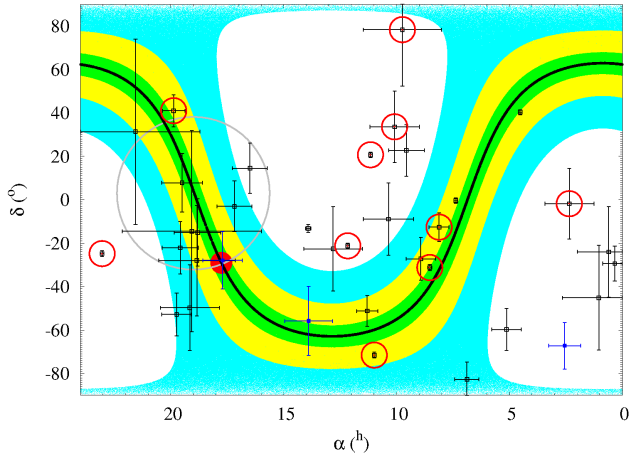}
       }
       \caption{Distribution in equatorial coordinates (right ascension and declination) of 36 events observed by IceCube. Data and errors 
                as in Supplementary Table 1 of Aartsen et al. (2014a). Events with deposited energies in excess of 1 PeV (3) appear in blue 
                (filled squares), the others in black (empty squares). Events containing muon tracks can be identified by their much 
                smaller positional error bars. The Galactic plane/equator appears as a continuous black line. The Galactic disk is 
                arbitrarily defined as the region confined between galactic latitude $-5$\degr and 5\degr (in green). The position of the 
                Galactic centre is represented by a filled red circle. The region enclosed between galactic latitude $-15$\degr and 15\degr 
                appears in yellow, and the one confined by $-30$\degr and 30\degr in cyan. See Table \ref{disk} for the number of events 
                found towards these regions. The large gray empty circle signals the approximate location of the Local Void. Event 18 is 
                the leftmost one, event 10 is the rightmost one; both are well away from the bulk of the Galactic disk. Red empty circles 
                signal probable background events (see the text for details).
               } 
       \label{data}
    \end{figure}

    The data used in this study appear in Supplementary Table 1 of Aartsen et al. (2014a). Thirty seven neutrino candidate events are 
    included there, but event 32 was produced by a coincident pair of background muons from unrelated air showers and that made it 
    impossible to reconstruct its arrival direction. This event will be excluded from our analysis as our initial working sample consists 
    of 36 events (see Fig. \ref{data}). Twenty-eight of them had shower-like topologies, while the remaining 8 had visible muon tracks. 
    Three events have deposited energies in excess of 1 PeV (in blue in Fig. \ref{data}). Event 14, with a deposited energy of nearly 1 PeV 
    (the third highest in the sample), is positionally coincidental (within the angular errors but almost 1\fdg5 from Sagittarius A$^{*}$) 
    with the location of the supermassive black hole of the Milky Way. Twenty-seven events have negative declination and, therefore, they 
    arrived from the southern hemisphere. This subsample of 27 events is a time- and deposited energy-limited, and perhaps complete, sample. 
%
%
%
%

    It is widely accepted that the observed neutrinos come from two different sources, atmospheric and extraterrestrial. Atmospheric 
    neutrinos are constantly being produced from all directions in the sky as they are the result of cosmic ray interactions in the upper 
    atmosphere and subsequent decay of mesons; their deposited energies are expected to be under 100 TeV. The neutrino candidate events in 
    Fig. \ref{data} correspond to a deviation from the atmospheric background at the 5.7$\sigma$ level (Aartsen et al. 2014a). This means 
    that out of 37 events, about 11 are background and the remaining 26 are extraterrestrial neutrinos. Aartsen et al. (2014a) already 
    pointed out that the properties of events 3, 8, 18, 28 and 32 are consistent with them being background detections.
%
%
    \begin{figure}
       \resizebox{\hsize}{!}{
          \includegraphics{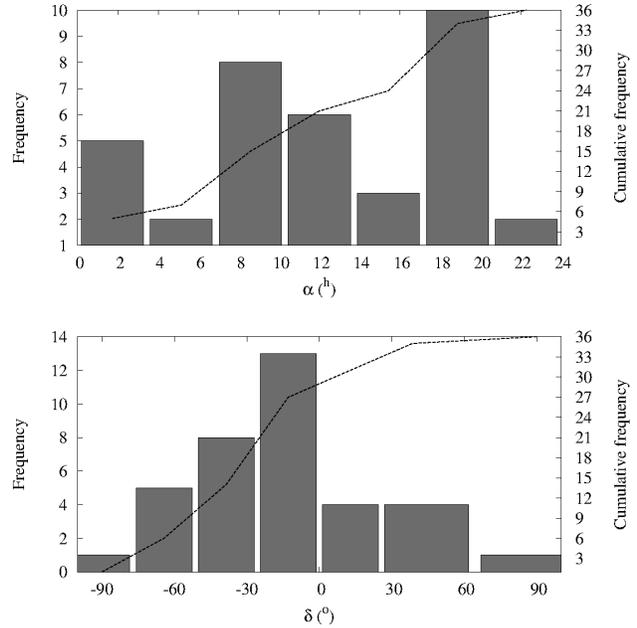}
       }
       \caption{Frequency distribution in equatorial coordinates (right ascension, top panel, and declination, bottom panel) of the 36 
                neutrino candidate events that appear in Supplementary Table 1 of Aartsen et al. (2014a). The number of bins is 
                2 $n^{1/3}$ where $n=36$. An obvious excess of events in the right ascension range 8$^{\rm h}$ to 20$^{\rm h}$ is observed.
               }
       \label{radec}
    \end{figure}
%
%

    Figure \ref{radec} shows the frequency distribution in equatorial coordinates ($\alpha, \delta$) of the 36 neutrino candidate events
    discussed above. There is an obvious anisotropy in declination induced by the fact pointed out above that 27 events arrived from the 
    southern celestial hemisphere. The source of this anisotropy is an observational bias, the IceCube neutrino observatory is more 
    sensitive to events arriving from that direction. There is also an obvious excess of events in the right ascension range 8$^{\rm h}$ to 
    20$^{\rm h}$. This range contains 27 events, outnumbering by 18 the 9 events in the ranges (0, 8)$^{\rm h}$ plus (20, 24)$^{\rm h}$; 
    this is a 4.2$\sigma$ departure from an isotropic distribution, where $\sigma=\sqrt{n}/2$ is the standard deviation for binomial 
    statistics and $n$ is the number of events. It may be argued that the identified excess of neutrino candidate events towards the right 
    ascension range 8$^{\rm h}$ to 20$^{\rm h}$ could be an effect of the instrumental acceptance. However, the Earth rotation guarantees 
    the uniform exposure of the detectors in right ascension (Aguilar et al. 2013). If we focus on the 27 southern events, there are 18
    of them within the right ascension range 8$^{\rm h}$ to 20$^{\rm h}$ and this still translates into a somewhat significant 2.45$\sigma$ 
    departure from an isotropic distribution. Besides, this excess of neutrino candidate events towards the right ascension range 8$^{\rm 
    h}$ to 20$^{\rm h}$ is consistent with the anisotropy in TeV cosmic rays identified by Schwadron et al. (2014). 

    In their study, Schwadron et al. (2014) have found a broad relative excess of cosmic rays between $\alpha \approx 8^{\rm h}$ and 
    $\alpha \approx 18^{\rm h}$. The details of the observed anisotropy depend on the cosmic ray energy range considered. The relative 
    excess in arrival direction is oriented towards the region of the sky that includes the downstream direction of the interstellar wind 
    and the downfield direction of the Local Interstellar Magnetic Field (LIMF; Schwadron et al. 2014). These TeV cosmic ray anisotropies 
    are related to the Interstellar Boundary Explorer (IBEX) ribbon (McComas et al. 2009). Neutrinos are not affected by the LIMF, but 
    high-energy cosmic ray particles are and they can produce neutrinos when they interact with atomic nuclei creating unstable particles 
    whose decay generates the neutrinos. As the parent high-energy cosmic ray particles can be deflected by magnetic fields, their original 
    sources are almost impossible to trace. In this scenario, it is theoretically possible to produce extraterrestrial neutrinos at 
    hundreds of AUs in the outskirts of the solar system; in fact, the structure observed by the IBEX spacecraft is a bright, narrow 
    ($\sim$20\degr wide) ribbon of energetic neutral atom emission that likely hints at the presence of an associated overdensity of 
    matter. A putative production of neutrinos towards the edge of the solar system could further complicate the interpretation of the 
    available data.

\begin{figure}
       \resizebox{\hsize}{!}{
          \includegraphics{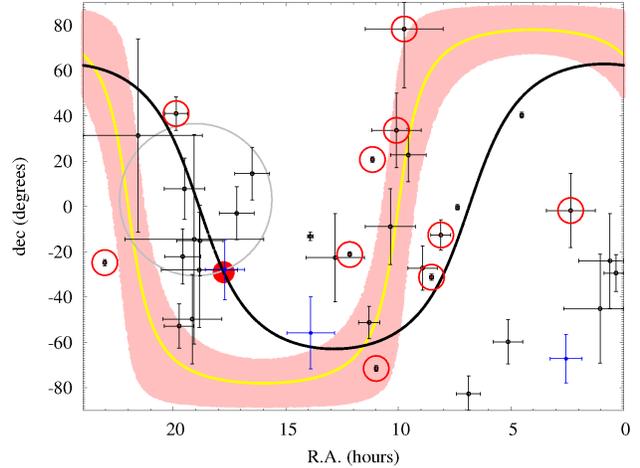}
       }
       \caption{Distribution in equatorial coordinates (right ascension and declination) of 36 events observed by IceCube as plotted in
                Fig. \ref{data}. The Galactic plane appears as a continuous black line. The position of the Galactic centre is represented 
                by a filled red circle. The large gray empty circle signals the approximate location of the Local Void. The continuous 
                yellow line follows the LISM magnetic equator as derived by Heerikhuisen et al. (2010). A narrow ribbon ($\sim$20{\degr} 
                wide) around the great circle defined by the LISM magnetic equator appears in pink. This area matches approximately the
                location of the IBEX ribbon (McComas et al. 2009).
               } 
       \label{dataibex}
    \end{figure}

    Figure \ref{dataibex} shows the location of the IBEX ribbon structure with respect to the Galactic equator and the IceCube events. The
    continuous yellow line follows the Local Interstellar Medium (LISM) magnetic equator as derived by Heerikhuisen et al. (2010). A 
    20{\degr} wide band around the LISM magnetic equator appears in pink. The IBEX ribbon is found within that area. Some events appear to 
    be aligned along the band. As the arrival distribution of high energy cosmic rays appears to be affected by perturbations induced in 
    the local interstellar magnetic field by the heliosphere wake (see, e.g., Desiati \& Lazarian 2012; Schwadron et al. 2014) some of the
    detected neutrinos may have been generated in that region.
%
%
%
%

    Returning to the connection between neutrino candidate events and discrete sources, Bai et al. (2014) have provided robust evidence on
    the possible Sagittarius A$^{*}$ origin of some of the events observed by IceCube. In particular, event 25 coincides spatially as well
    as temporally with some transient activity observed in X-rays towards Sagittarius A$^{*}$ on 2012 February 9. Another notable spatial
    coincidence is the galaxy 2dFGRS TGS070Z303 and event 18 (a muon-track event), separated by just over 59 arcseconds. Galaxy 2dFGRS 
    TGS070Z303 has a radial velocity of 37,866 km s$^{-1}$ (Colless et al. 2001). Also the galaxy 2dFGRS TGS280Z153 and event 10 (of 
    shower-like topology) are separated by 2\farcm18. Galaxy 2dFGRS TGS280Z153 has a radial velocity of 53,795 km s$^{-1}$ (Colless et al. 
    2001). An additional, perhaps significant, coincidence is the LINER AGN 2MASX J13550744-1309568 that is 5\farcm16 from event 23, 
    another muon-track event at coordinates $\alpha$ = 13\fh91, $\delta = -$13\fdg2. AGN 2MASX J13550744-1309568 has a radial velocity of 
    24,924~km~s$^{-1}$ (Mauch \& Sadler 2007).

 \section{A Monte Carlo-based angular distribution analysis}
    Let us consider a sample of $n$ events observed projected on the surface of a sphere, the sky in our case. We want to study the 
    possible non-uniform spatial distribution of these events. In order to do that, we generate random points on the surface of a sphere 
    using an algorithm due to Marsaglia (1972). The positions of these points on the sky have right ascension in the range (0$^{\rm h}$, 
    24$^{\rm h}$) and declination in the range (-90\degr, 90\degr). We perform 10$^{7}$ experiments, each one including $n$ events, 
    randomly distributed on the sky. In our histograms, the adopted number of bins is 2 $n_{\rm p}^{1/3}$, where $n_{\rm p}$ is the number 
    of pairs. We compute the angular distance, $d$, between two given points (or events) using the equation $\cos d = \cos(\alpha_1 - 
    \alpha_2) \cos\delta_1 \cos\delta_2 + \sin\delta_1 \sin\delta_2$, where $(\alpha_1, \delta_1)$ and $(\alpha_2, \delta_2)$ are the 
    equatorial coordinates of both points. Then count the number of pairs per bin and divide by the number of experiments performed. 

\begin{figure}
       \resizebox{\hsize}{!}{
          \includegraphics{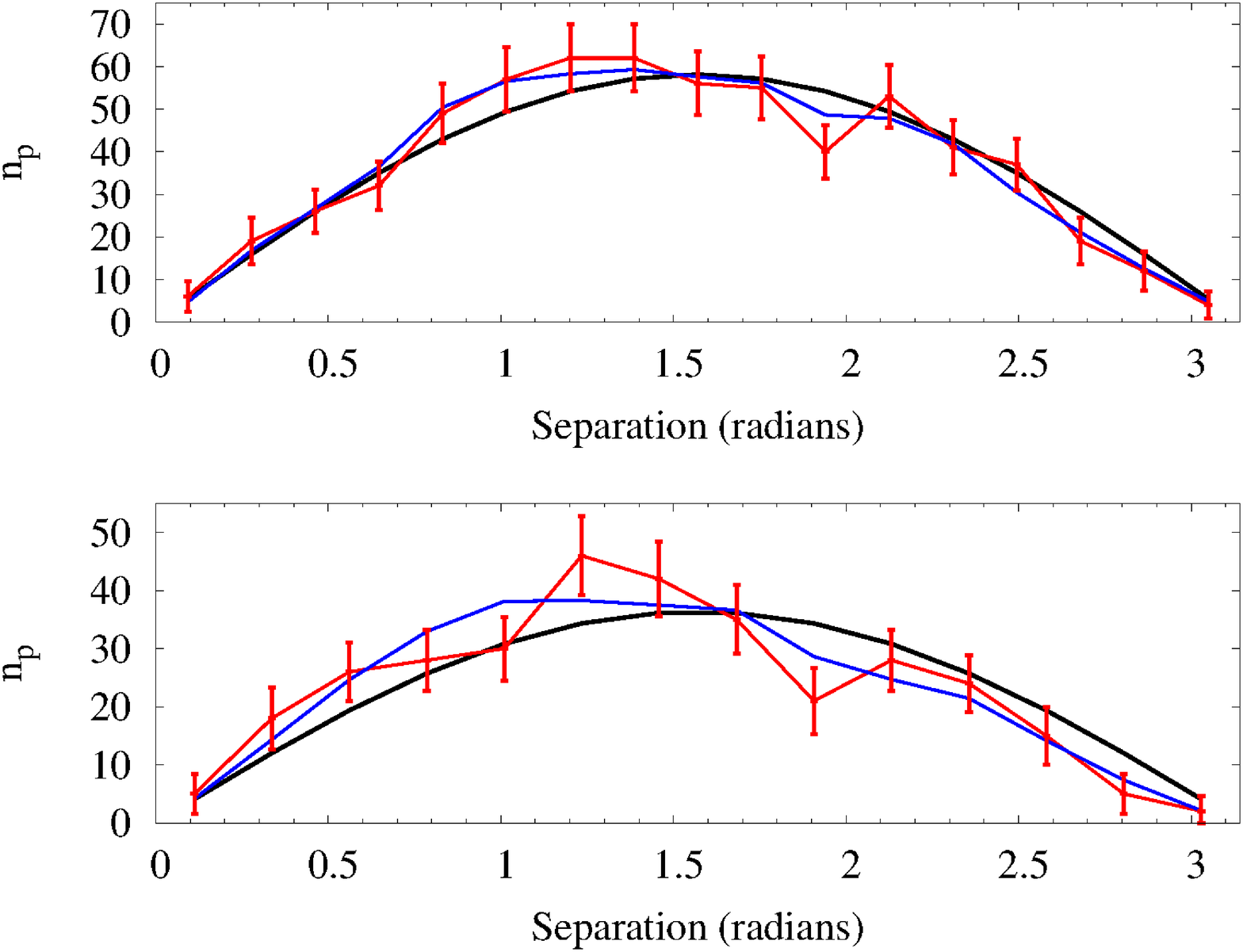}
       }
       \caption{Separation distribution for the events in Supplementary Table 1 of Aartsen et al. (2014a), red line/points with error bars, 
                the number of pairs is $n_{\rm p}$. The theoretical separation distribution for an uniformly spread sample across the sky 
                of $n$ = 36 events (or 630 pairs, top panel) and $n$ = 26 events (or 325 pairs, bottom panel) is displayed as a continuous 
                black line. The top panel displays the entire sample of 36 events and the bottom panel shows the results excluding events 
                of probable terrestrial origin, a subsample of 26. The blue lines correspond to the average separation distribution when 
                the positional errors associated with the events are taken into account (see the text for details). The 1$\sigma$ error 
                bars have been computed as described in the text. The number of bins is 2 $n_{\rm p}^{1/3}$.
               }
       \label{iso}
    \end{figure}

    The black continuous curves in Fig. \ref{iso} show the expected distribution of separations between events on the sky if they are 
    uniformly distributed. The most probable separation is $\pi/2$ radians or 90\degr. The distribution associated with the data in 
    Supplementary Table 1 of Aartsen et al. (2014a) is far from regular (top panel) but only one bin deviates $\sim$2$\sigma$ 
    (2.25$\sigma$) from a uniform distribution. Here, we use the approximation given by Gehrels (1986) when $n_{\rm p} < 21$ to compute the 
    error bars as Poisson statistics applies: $\sigma \sim 1 + \sqrt{0.75 + n_{\rm p}}$ ($\sigma = \sqrt{n_{\rm p}}$ otherwise), where 
    $n_{\rm p}$ is the number of pairs. In principle, all the events are uncorrelated and distributed according to Poisson statistics. The 
    addition of two independent Poissonian variables (considering terrestrial and extraterrestrial events, see above) is still Poissonian.

    Figure \ref{iso}, top panel, does not take into account the fact that a fraction of the events (11 out of 37) likely have terrestrial 
    origin. Only the extraterrestrial neutrinos are of interest here, i.e. the sample must be decontaminated. Background rejection is 
    implemented following the conclusions obtained by Aartsen et al. (2014a), and also using an algorithm in which we assume that they will 
    have deposited energies under 100 TeV and that (approximately and based on Fig. 2 in Aartsen et al. 2014a) 8 background events have 
    energies in the range 30--53.3 TeV, 2 in the range 53.3--76.7 TeV, and 1 in the range 76.7--100 TeV. In addition, we rank events based 
    on how high is their probability of being part of an isotropic distribution. At the energies of interest here, the angular distribution 
    of atmospheric neutrinos is expected to be very nearly isotropic. We compute the separation distributions for all the possible 
    combinations (more than 600 million) of 11 events and compare with that of an isotropic distribution of the same size. Those 
    combinations of 11 events with root-mean-square deviation from the isotropic behaviour $<$ 0.0047 are selected and the number of times 
    each event appears is counted. Statistically speaking, those with the highest number of counts are more likely to be part of an 
    isotropically distributed subsample. Using the conclusions from Aartsen et al. (2014a) and the events ranked as described, we conclude 
    that the most probable background events are: 1, 3, 8, 9, 18, 27, 28, 29, 31, 32 and 37. These events appear as red empty circles in 
    Fig. \ref{data} (for a different approach to the issue of background rejection see Padovani \& Resconi 2014).
     
    If we focus on the 26 events (or 325 pairs) which are most probably of extraterrestrial origin, we obtain the bottom panel in Fig. 
    \ref{iso}. For this decontaminated sample, the difference between the theoretical (isotropic) value and the actual one is always $< 
    2.5\sigma$. However, the range 0 to $\pi$/2 radians contains 195 pairs, outnumbering by 65 the 130 pairs in the range $\pi$/2 to $\pi$ 
    radians; this is a 3.6$\sigma$ departure from an isotropic distribution, where $\sigma=\sqrt{n_{\rm p}}/2$ is now the standard 
    deviation for binomial statistics. Therefore, the irregular distribution on the celestial sphere is different from that of a sample of 
    points uniformly scattered, i.e. an anisotropy exists. An excess of events appears to be located towards the equatorial coordinates 
    (18\fh7, $-23$\degr); this has already been pointed out by Aartsen et al. (2013c). This region is relatively far from the Galactic disk, 
    the densest part of the Milky Way Galaxy, and towards the Local Void (Tully \& Fisher 1987). This area of the sky has a confirmed low 
    local density of galaxies with radial velocities below 3,000 km s$^{-1}$ (see, e.g., Fig. 2 in Nasonova \& Karachentsev 2011). This 
    region of low density extends up to distances of 40 to 60 Mpc centred at least 23 Mpc from the Milky Way (Tully et al. 2008) and it may 
    be even larger, perhaps connecting with the Microscopium/Sagittarius void (Kraan-Korteweg et al. 2008). The geometric centre of the 
    Local Void is found towards the equatorial coordinates (19\fh0, +3\degr), starting at the edge of the Local Group (Nasonova \& 
    Karachentsev 2011). The large gray empty circle in Fig. \ref{data} signals the approximate location and projected extension of the Local 
    Void. Taking into account the error bars, it encloses nearly 25\% of all the IceCube events.
%
%
%
%

    On the other hand, the Galactic disk is the brightest source of gamma rays in the sky, can the excess of events pointed out above have 
    their origin in the disk? To test this hypothesis we count the number of IceCube events within bands whose boundaries are limited in 
    galactic latitude, $b$ (see Table \ref{disk}). For example, points within the Galactic disk have $|\Delta b|$ = 5\degr and are confined 
    between galactic latitude $-5$\degr and 5\degr (green area in Fig. \ref{data}). We repeat the process using the same uniformly 
    distributed model studied before. Differences between the observational data and the isotropic model are not statistically significant. 
    It appears unlikely that the dominant source of these high-energy neutrinos is in or near the Galactic disk. This is somewhat 
    consistent with results obtained by Aartsen et al. (2014a); in their work, it is found that the correlation with the Galactic plane is
    not significant, with a probability of 2.8\% for a width of the disk of $\pm$7\fdg5. Restricting the analysis to the range of right 
    ascension 16$^{\rm h}$ to 22$^{\rm h}$ does not show any significant changes and that also makes the Galactic bulge an unlikely source. 
    Results are equivalent for the full sample and the decontaminated one (see Table \ref{disk}, in boldface). 
     
%
%
    \begin{table}
       \centering
       \fontsize{8}{11pt}\selectfont
       \tabcolsep 0.15truecm
       \caption{Number of events detected by IceCube as a function of the galactic latitude, $b$, or distance from the Galactic plane in 
                steps of 5\degr and the expected number of events derived from an isotropic model (see the text) with its associated error. 
                The difference between the observational data and the isotropic model in units of $\sigma$ is also given. Results for the 
                decontaminated sample appear in boldface. See Fig. \ref{data} for a reference on the various $|\Delta b|$ regions.
               }
       \begin{tabular}{cccc}
          \hline
           $|\Delta b|$ (\degr) & isotropic ($\pm\sigma$) & IceCube data & Difference (in units of $\sigma$) \\
          \hline
           5                    &  3.14$\pm$2.97          & 3            & 0.05                              \\
                                & {\bf  2.27$\pm$2.74}    & {\bf  2}     & {\bf 0.10}                        \\
           10                   &  6.25$\pm$3.65          & 10           & 1.03                              \\
                                & {\bf  4.51$\pm$3.29}    & {\bf  8}     & {\bf 1.06}                        \\
           15                   &  9.32$\pm$4.17          & 15           & 1.36                              \\
                                & {\bf  6.73$\pm$3.73}    & {\bf 11}     & {\bf 1.14}                        \\
           20                   & 12.31$\pm$4.61          & 16           & 0.80                              \\
                                & {\bf  8.89$\pm$4.10}    & {\bf 12}     & {\bf 0.75}                        \\
           25                   & 15.21$\pm$4.99          & 18           & 0.56                              \\
                                & {\bf 10.98$\pm$4.42}    & {\bf 14}     & {\bf 0.68}                        \\
           30                   & 18.00$\pm$5.33          & 20           & 0.38                              \\
                                & {\bf 13.00$\pm$4.71}    & {\bf 16}     & {\bf 0.64}                              \\
         \hline
       \end{tabular}
       \label{disk}
    \end{table}
%
%

    Our previous analysis has not taken into consideration that the positional errors associated with the events, in particular to those of 
    shower topology, are very large, much larger than those typical of optical astrometry. What is the impact of these errors on our 
    previous analysis? In order to evaluate the effect of large angular errors on the separation distribution, we generate sets of 
    synthetic events Monte Carlo-style (Metropolis \& Ulam 1949) using the equatorial coordinates and median angular errors in 
    Supplementary Table 1 of Aartsen et al. (2014a). For a given event of coordinates $(\alpha, \delta)$ and median angular error 
    $\Delta e$, we generate a synthetic event of coordinates $(\alpha_{\rm s}, \delta_{\rm s})$ where the new coordinates are such that the
    angular distance between the synthetic coordinates and the original ones is given by $d^{*} = |{\rm GRND}| \ \Delta e$, with GRND being 
    a random number in the interval (-1, 1) following a Gaussian distribution obtained using the Box-Muller method (Press et al. 2007) with 
    zero mean and unit variance. If the North Celestial Pole is within the error circle, declination is restricted to the range $\delta -
    \Delta e$ to $+\pi$/2, when the South Celestial Pole is included the declination range is $-\pi$/2 to $\delta + \Delta e$; in both 
    cases, the right ascension can take any value in the range 0$^{\rm h}$ to 24$^{\rm h}$ as parts of all meridians are within the error 
    circle as well. This approach is not singular at the poles. When none of the poles is included within the error circle, the range in 
    declination is $\delta - \Delta e$ to $\delta + \Delta e$ and the equivalent in right ascension is $\alpha - \Delta e$ to $\alpha + 
    \Delta e$. The average separation distribution for 10$^{7}$ sets of synthetic events based on the original ones and their positional 
    errors is displayed as a blue line in Fig. \ref{iso} (top panel). Taking into account the positional errors significantly reduces the 
    difference between the observational angular separation distribution and that of an isotropic case. However, for the clean sample 
    (bottom panel) there is still a 3.1$\sigma$ departure from an isotropic distribution when considering separations smaller than $\pi$/2. 
    The approach outlined here has also been used to strengthen the results of our decontamination algorithm (see above) and our list of 
    most probable background events has been obtained after hundreds of trials that use the median angular errors.

 \section{The atmospheric background}
\begin{figure}
       \resizebox{\hsize}{!}{
          \includegraphics{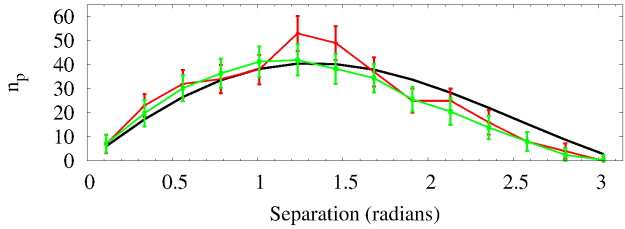}
       }
       \caption{Separation distribution (red line/points with error bars, top panel) for the 27 events located towards the southern
                hemisphere (negative declination) in Supplementary Table 1 of Aartsen et al. (2014a), the number of pairs is $n_{\rm p}$. 
                The analogous separation distribution for an uniformly distributed sample across a semisphere of $n$ = 27 events or 531
                pairs is displayed as a continuous black line. The continuous green line shows the average of $10^7$ experiments where the
                right ascension is randomized as described in the text.
               }
       \label{shuffle}
    \end{figure}

    The background produced by atmospheric neutrinos is of diffuse nature. These atmospheric neutrinos result from cosmic rays interacting
    with the atmosphere. So far, our analysis has assumed that this background follows a mathematically isotropic model as the theory 
    predicts. However, in the calculations carried out by the IceCube collaboration, the significance of their results is evaluated by 
    comparing with a distribution obtained by performing yet another analysis over a sample of only-background data sets that are obtained 
    by scrambling real data in right ascension. In other words, right ascension is randomized while keeping the original values in 
    declination. This is done to take into account that the detector has anisotropic acceptance, i.e. its instrumental response function 
    is unknown. This approach assumes that the unknown instrumental response function does not change over time, i.e. it remains the same
    over the entire search period. Figure \ref{shuffle} shows that under this approach the separation distribution of the 27 southern 
    events and that of an average of $10^7$ only-background data sets are compatible, i.e. the angular distribution of the events is 
    isotropic. However, deviations close to 1.5$\sigma$ are observed.
%
%
%
%

 \section{Discussion}
    After being emitted, neutrinos move unaffected by gas, dust or magnetic fields. The somewhat significant excess observed towards the 
    equatorial coordinates (18\fh7, $-23$\degr) is an important clue to the origin of the observed events as their associated 
    neutrinos have travelled very little distance (if their sources are Galactic) or across mostly empty space in their final leg, the 
    Local Void (if they are of extragalactic origin). This is probably the key fact to understand the place of origin of the majority of 
    events detected so far by the IceCube neutrino observatory, extragalactic versus local. Our analysis shows no statistically significant 
    excess of events associated with the Galactic disk (see above). If sources in the Galactic halo are expected to follow a uniform 
    spread, then our findings are somewhat incompatible with such origin. As for sources within the Local Group, multiple events are 
    observed towards an area of the sky where the density of galaxies with radial velocities below 3,000 km s$^{-1}$ is quite low. The 
    alternative explanation is straightforward: sources located at cosmological distances, well beyond the Local Group. 

\begin{figure}
       \resizebox{\hsize}{!}{
          \includegraphics{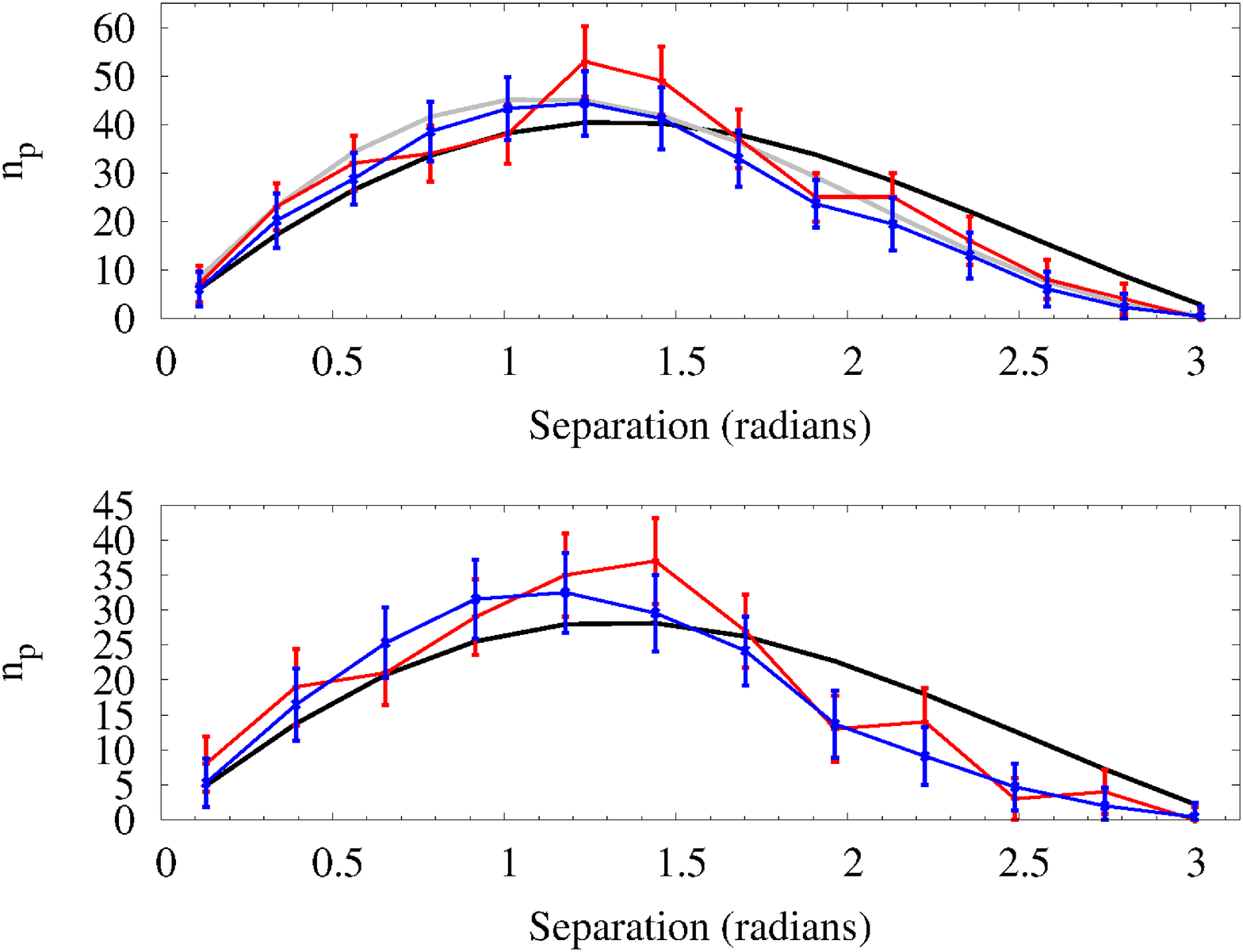}
       }
       \caption{Separation distribution (red line/points with error bars, top panel) for the 27 events located towards the southern 
                hemisphere (negative declination) in Supplementary Table 1 of Aartsen et al. (2014a), the number of pairs is $n_{\rm p}$. 
                The analogous separation distribution for an uniformly distributed sample across a semisphere of $n$ = 27 events or 531 
                pairs is displayed as a continuous black line in both panels. The average separation distribution for a set of synthetic 
                events obtained from the original events and their Gaussian positional errors appears as a blue line, the average number of 
                synthetic events is $\sim$25.8 and the associated number of pairs is $\sim$319.9, more details in the text. The gray line 
                shows the best fractal-like distribution (see the text for details). The bottom panel plots the equivalent results for the
                decontaminated sample of 21 events. The 1$\sigma$ error bars have been computed as described in the text. The number of 
                bins is 2 $n_{\rm p}^{1/3}$, where $n_{\rm p}$ is 531 (top) and 210 (bottom). 
               }
       \label{fractal}
    \end{figure}

    The spatial distribution of galaxies situated at cosmological distances has been found to be hierarchical (see, e.g., Pietronero 1987) 
    and this is also consistent with numerical expectations since early studies (e.g. Aarseth et al. 1979). In an attempt to 
    further explore this interesting possibility, a dominant extragalactic origin for these events, we restrict our analysis to the sample 
    of 27 events located towards the southern hemisphere in Supplementary Table 1 of Aartsen et al. (2014a). The separation distribution of 
    this subsample (on the surface of a semisphere) is quite different from that in Fig. \ref{iso}, with a reduced number of pairs 
    separated by more than $\pi/2$ radians (the average angular distance for points uniformly spread on a sphere) with respect to the full 
    sample (see Fig. \ref{fractal}). 

    In order to check if the angular distribution of events on the semisphere is isotropic, we proceed as we did with the full sample and 
    perform 10$^{7}$ experiments, each one including 27 events randomly distributed on the southern half of the celestial sphere, counting 
    the number of pairs within a certain range of separations. The results are plotted as a black continuous line in Fig. \ref{fractal}. 
    Differences between the uniform case and the subsample are consistent across the entire separation range and they are marginally 
    significant for some bins (the largest difference is 1.8$\sigma$). Our analysis shows that, for events observed towards the southern 
    hemisphere, the angular distribution is not isotropic. There is a 3.7$\sigma$ departure from an isotropic distribution when considering 
    separations smaller than $\pi$/2; as pointed out above, the deviation for the entire celestial sphere is 3.6$\sigma$. If we replace the 
    original set of 27 data by sets of synthetic events obtained from the original ones and their Gaussian positional errors (see above) 
    and average the results of 10$^{7}$ experiments, we obtain the blue curve in Fig. \ref{fractal} (top panel) that deviates even more 
    clearly from the isotropic arrival scenario (black curve) for separations larger than the average (the largest difference is now close 
    to 3$\sigma$ near the high end of the distribution). We observe 98 pairs with angular separation in the range $\pi$/2 to $\pi$ instead
    of the expected 149 in the isotropic case, a 5.5$\sigma$ deviation. This reduced number of pairs characterized by large separations 
    strongly favours the existence of some type of weak but noticeable clustering for these neutrino sources. Absence of strong anisotropy 
    is usually interpreted as an indication of cosmological origin. Using the decontaminated sample (see Fig. \ref{fractal}, bottom panel), 
    we observe a 3.9$\sigma$ departure (149 observed pairs instead of 121 in the isotropic case for a total of 210 pairs) from an isotropic 
    distribution when considering separations smaller than $\pi$/2. When taking into account the positional errors the statistical 
    significance is reduced to 2.8$\sigma$.
%
%
%
%

    If the sources of these neutrinos are located at cosmological distances, probably within remote galaxy clusters, one may expect that 
    they are organized following a hierarchical fractal model. Star formation within the Milky Way follows a fractal pattern (e.g. de la 
    Fuente Marcos \& de la Fuente Marcos 2006a); similar patterns are observed in starburst galaxies (e.g. de la Fuente Marcos \& de la 
    Fuente Marcos 2006b,c). Soneira \& Peebles (1978) showed that the appearance of the projected distribution of galaxies can be 
    reproduced using a model based on the superposition of fractal clumps. In order to test this hipothesis we consider $n_{\rm a}$ 
    uniformly distributed points, $(\alpha_n, \delta_n)$, on the surface of the unit semisphere. Each point is the centre of an arc of 
    radius $R/\lambda$. Within each arc we consider $n_{\rm a}$ additional uniformly distributed points, each one is centre for an 
    associated arc of radius $R/\lambda^{2}$. Within each arc and on the surface of the unit semisphere, we generate $n_{\rm a}$ uniformly 
    distributed points. The approach is similar to the one described by Soneira \& Peebles (1978). 

    In our case, and in order to have 27 points (events) per experiment, we consider $n_{\rm a}$ = 3, $R$ in the range $\sim$0 to $\sim2\pi$ 
    and $\lambda \in (0, 2)$. After performing a large number of tests, each one consisting of 10$^{6}$ experiments, the best fit to the 
    southern hemisphere data is found for $R = \pi$ and $\lambda = 0.9$. The gray curve in Fig. \ref{fractal}, top panel, shows the 
    distribution of separations for our fractal-like model. Although unable to reproduce the asymmetric behaviour at separations in the 
    range 0.5-1.5 radians (top panel), its deviations from the actual data are not statistically significant and its overall agreement 
    improves for the average separation distribution of synthetic events (blue curve in Fig. \ref{fractal}, top panel). These deviations are 
    always smaller than those from a uniform distribution. In order to find out how significant the differences between the two models 
    (fractal-like versus isotropic) are and because the sample is rather small, we use the likelihood formula due to Saha (1998). The 
    average value of the $W$ estimator for the fractal-like model is 6 times greater than the one for the isotropic model; therefore, it is 
    statistically more robust. The distribution of events over the southern hemisphere is almost certainly anisotropic. 
     
 \section{Conclusions}
    In this paper, we have studied the angular distribution of IceCube high-energy events trying to answer the astrophysically relevant 
    question of whether or not their spread is isotropic to conclude that the observed events are very probably not evenly distributed on 
    the sky. Differences between a uniform angular distribution and the observed one are indeed marginally significant at the 2$\sigma$ 
    level or better. Our main findings can be summarized as follows:
    \begin{itemize}
       \item The angular distribution of IceCube high-energy events exhibits a modest but statistically significant anisotropy.
       \item There is no statistically significant excess of events projected towards the Galactic disk.
       \item The observed distribution of events appears to be incompatible with an origin in the Milky Way halo.
       \item A relatively significant excess of events appears projected towards the Local Void. This suggests either a Galactic origin or
             one well beyond the Local Group for these events.
       \item The angular distribution of events towards the southern hemisphere appears to be anisotropic and it is statistically more 
             compatible with an origin in a distant, fractal-like organized structure than with the currently assumed isotropicity. 
    \end{itemize}
    We must stress that our results are based on small number statistics. However, they are compatible with well-established results for 
    GRBs, namely that short GRBs are distributed anisotropically (Bal\'azs et al. 1998). Anisotropy is also observed for the highest energy 
    cosmic rays (e.g. Abreu et al. 2012; Schwadron et al. 2014). Because of the small size of the current data set compared to those of 
    short GRBs or high-energy cosmic rays, the results of our study are necessarily limited in scope as only very large effects can be 
    reliably identified. Although a small fraction of closer sources is possible (perhaps those putatively related to the IBEX ribbon), 
    remote hierarchical structures appear to be the main source of these energetic neutrinos.

 \begin{acknowledgements}
    In preparation of this paper, we made use of the NASA Astrophysics Data System and the ASTRO-PH e-print server. This research has made 
    use of the SIMBAD database and the VizieR service, operated at CDS, Strasbourg, France.
 \end{acknowledgements}


\begin{thebibliography}{ }
    \bibitem{AA79} Aarseth, S.J., Turner, E.L., \& Gott, J.R. III 1979,
                   ApJ, 228, 664
    \bibitem{A13a} Aartsen, M.G., Abbasi, R., Abdou, Y., et~al. (IceCube Coll.) 2013a,
                   Sci, 342, 1242856
    \bibitem{A13b} Aartsen, M.G., Abbasi, R., Abdou, Y., et~al. (IceCube Coll.) 2013b,
                   Phys. Rev. Lett., 111, 021103
    \bibitem{A13c} Aartsen, M.G., Abbasi, R., Ackermann, M., et~al. (IceCube Coll.) 2013c,
                   Phys. Rev. D, 88, 112008
    \bibitem{A14a} Aartsen, M.G., Ackermann, M., Adams, J., et~al. (IceCube Coll.) 2014a,
                   Phys. Rev. Lett., 113, 101101
    \bibitem{A14b} Aartsen, M.G., Ackermann, M., Adams, J., et~al. (IceCube Coll.) 2014b,
                   ApJ, 796, 109
    \bibitem{AB12} Abreu, P., Aglietta, M., Ahlers, M., et~al. 2012,
                   J. Cosmol. Astropart. Phys., 4, 40 
    \bibitem{AG13} Aguilar, J.A., Aartsen, M.G., Abbasi, R., et~al. (IceCube Coll.) 2013,
                   Nuclear Physics B - Proceedings Supplements, 239--240, 184 (arXiv:1010.6263)
    \bibitem{AM14} Ahlers, M., \& Murase, K. 2014,
                   Phys. Rev. D, 90, 023010
    \bibitem{N14b} Anchordoqui, L.A., Goldberg, H., Lynch, M.H., et~al. 2014a,
                   Phys. Rev. D, 89, 083003
    \bibitem{N14a} Anchordoqui, L.A., Paul, T.C., da Silva, L.H.M., Torres, D.F., \& Vlcek, B.J. 2014b,
                   Phys. Rev. D, 89, 127304 
    \bibitem{AR89} Arnett, W.D., Bahcall, J.N., Kirshner, R.P., \& Woosley, S.E. 1989,
                   ARA\&A, 27, 629 
    \bibitem{BA05} Bahcall, J.N. 2005,
                   Phys. Scr., T121, 46
    \bibitem{BA14} Bai, Y., Barger, A.J., Barger, V., et~al. 2014,
                   Phys. Rev. D, 90, 063012
    \bibitem{BA98} Bal\'azs, L.G., M\'esz\'aros, A., \& Horv\'ath, I. 1998,
                   A\&A, 339, 1
    \bibitem{BE14} Becker Tjus, J., Eichmann, B., Halzen, F., Kheirandish, A., \& Saba, S.M. 2014,
                   Phys. Rev. D, 89, 123005
    \bibitem{BE04} Bellerive, A. 2004,
                   Int. J. Mod. Phys. A, 19, 1167
    \bibitem{BH10} Bhattacharya, A., Choubey, S., Gandhi, R., \& Watanabe, A. 2010,
                   J. Cosmol. Astropart. Phys., 9, 9
    \bibitem{CH14} Chen, C.-Y., Dev, P.S.B., \& Soni, A. 2014,
                   Phys. Rev. D, 89, 033012
    \bibitem{CO01} Colless, M., Dalton, G., Maddox, S., et al., 2001,
                   MNRAS, 328, 1039
    \bibitem{M06a} de la Fuente Marcos, R., \& de la Fuente Marcos, C. 2006a,
                   A\&A, 452, 163
    \bibitem{M06b} de la Fuente Marcos, R., \& de la Fuente Marcos, C. 2006b,
                   A\&A, 454, 473
    \bibitem{M06c} de la Fuente Marcos, R., \& de la Fuente Marcos, C. 2006c,
                   MNRAS, 372, 279
    \bibitem{DL12} Desiati, P., \& Lazarian, A. 2012,
                   ApJ, 762, 44
    \bibitem{EM14} Ema, Y., Jinno, R., \& Moroi, T. 2014,
                   Phys. Lett. B, 733, 120
    \bibitem{GE86} Gehrels, N. 1986,
                   ApJ, 303, 336
    \bibitem{HA14} Halzen, F. 2014,
                   Astron. Nachr., 335, 507
    \bibitem{HE10} Heerikhuisen, J., Pogorelov, N.V., Zank, G.P., et~al. 2010,
                   ApJ, 708, L126
    \bibitem{JO14} Joshi, J.C., Winter, W., \& Gupta, N. 2014,
                   MNRAS, 439, 3414
    \bibitem{KK08} Kraan-Korteweg, R.C., Shafi, N., Koribalski, B.S., et~al. 2008,
                   Outlining the Local Void with the Parkes HI ZOA and Galactic Bulge Surveys,
                   in Galaxies in the Local Universe, ed. B.S. Koribalski, \& H. Jerjen, 
                   Springer Netherlands, Astrophysics and Space Science Proceedings, 13  
    \bibitem{KR14} Krau\ss\, F., Kadler, M., Mannheim, K., et~al. 2014,
                   A\&A, 566, L7 
    \bibitem{LI14} Liu, R.-Y., Wang, X.-Y., Inoue, S., Crocker, R., \& Aharonian, F. 2014,
                   Phys. Rev. D, 89, 083004
    \bibitem{LU14} Lunardini, C., Razzaque, S., Theodoseau, K.T., \& Yang, L. 2014,
                   Phys. Rev. D, 90, 023016
    \bibitem{MA72} Marsaglia, G. 1972,
                   Ann. Math. Stat., 43, 645
    \bibitem{MS07} Mauch, T., \& Sadler, E.M. 2007,
                   MNRAS, 375, 931
    \bibitem{MC09} McComas, D.J., Allegrini, F., Bochsler, P., et~al. 2009,
                   Sci, 326, 959
    \bibitem{MU49} Metropolis, N., \& Ulam, S. 1949,
                   J. Am. Stat. Assoc., 44, 335
    \bibitem{MU13} Murase, K., Ahlers, M., \& Lacki, B.C. 2013,
                   Phys. Rev. D, 88, 121301
    \bibitem{NK11} Nasonova, O.G., \& Karachentsev, I.D. 2011,
                   Astrophys., 54, 1 
    \bibitem{PR14} Padovani, P., \& Resconi, E. 2014,
                   MNRAS, 443, 474 
    \bibitem{PI87} Pietronero, L. 1987,
                   Physica A, 144, 257
    \bibitem{PR07} Press, W.H., Teukolsky, S.A., Vetterling, W.T., \& Flannery, B.P. 2007,
                   Numerical Recipes: The Art of Scientific Computing, 3rd Edition
                   (Cambridge Univ. Press, Cambridge), 364 
    \bibitem{R13a} Razzaque, S. 2013a,
                   Phys. Rev. D, 88, 081302
    \bibitem{R13b} Razzaque, S. 2013b,
                   Phys. Rev. D, 88, 103003
    \bibitem{SA98} Saha, P. 1998,
                   AJ, 115, 1206
    \bibitem{SC14} Schwadron, N.A., Adams, F.C., Christian, E.R., et~al. 2014,
                   Sci, 343, 988
    \bibitem{SP78} Soneira, R.M., \& Peebles, P.J.E. 1978,
                   AJ, 83, 845
    \bibitem{TA14} Taylor, A.M., Gabici, S., \& Aharonian, F. 2014,
                   Phys. Rev. D, 89, 103003
    \bibitem{TF87} Tully, R.B., \& Fisher, J.R. 1987,
                   Nearby Galaxies Atlas
                   (Cambridge Univ. Press, Cambridge)
    \bibitem{TU08} Tully, R.B., Shaya, E.J., Karachentsev, I.D., et~al. 2008,
                   ApJ, 676, 184 
    \bibitem{WI13} Winter, W. 2013,
                   Phys. Rev. D, 88, 083007
    \bibitem{ZA14} Zavala, J. 2014,
                   Phys. Rev. D, 89, 123516
 \end{thebibliography}
\end{document}